# Access to Emergency Services: A New York City Case Study

**Authors**:

Sukhwan Chung[a], Madison Smith[a], Andrew Jin[a], Luke Hogewood[a], Maksim Kitsak[b], Jeffrey Cegan[a], Igor Linkov[a]

**Affiliations**:

[a]Risk and Decision Science Team, US Army Engineer Research and Development Center – Environmental Laboratory, US Army Corps of Engineers
696 Virginia Rd, Concord, MA, 01742-2718, USA

[b]Faculty of Electrical Engineering, Mathematics and Computer Science, Delft University of Technology
Van Mourik Broekmanweg 6, 2628 XE, Delft, The Netherlands

## Abstract

Emergency services play a crucial role in safeguarding human life and property within society. In this paper, we propose a network-based methodology for calculating transportation access between emergency services and the broader community. Using New York City as a case study, this study identifies 'emergency service deserts' based on the National Fire Protection Association (NFPA) guidelines, where accessibility to Fire, Emergency Medical Services, Police, and Hospitals are compromised. The results show that while 95% of NYC residents are well-served by emergency services, the residents of Staten Island are disproportionately underserved. By quantifying the relationship between first responder travel time, Emergency Services Sector (ESS) site density, and population density, we discovered a negative power law relationship between travel time and ESS site density. This relationship can be used directly by policymakers to determine which parts of a community would benefit the most from providing new ESS locations. Furthermore, this methodology can be used to quantify the resilience of emergency service infrastructure by observing changes in accessibility in communities facing threats.

# 1. Introduction

Emergency services are an important aspect of the critical functions of a society that safeguard human life and property. According to the Department of Homeland Security (DHS) Cybersecurity & Infrastructure Security Agency (CISA), the Emergency Services Sector (ESS) is a community of trained personnel and resources, including Fire and Rescue Services (Fire), Emergency Medical Services (EMS), Law Enforcement (Police), Emergency Management, and Public Works, that provide prevention, preparedness, response, and recovery services during both day-to-day operations and incident response [CISA 2023]. The effectiveness of these services often relies on their accessibility to provide quick responses to the general public during emergencies.

Numerous studies have corroborated the correlation between response times and the effectiveness of emergency services delivered by Police [Vidal 2018], Hospitals and Emergency Medical Services (EMS) [Blackwell 2002, Pons 2005, Jaldell 2014], and Fire Services [Challands 2010, Jaldell 2017]. All of these studies found that faster response times lead to desirable outcomes in emergency systems. For example, faster police response increases the likelihood of clearing crimes, faster paramedic response improves the survival benefits of patients, and faster firefighter response reduces the cost of structural damages. Given this strong motivation to reduce response times of first responders and increase general public's access to emergency services, it is essential to develop methods to measure accessibility based on the response times of existing emergency services.

There are various methods that exist for measuring first responder response times required for a safer community. Due to the abundance of literature on methodology, we have chosen to focus specifically on studies that analyzed accessibility to emergency services by using first responder travel times on road networks. Many of the studies focused on access to a single emergency system, often the healthcare system (such as emergency rooms, hospitals, clinics, or other health institutions) [e.g., Carr 2009, Schuster 2024]. However, access to Fire & Rescue, and Emergency Medical Service [e.g., Rohr 2020, Jeon 2018] or Police [DeAngelo 2023, Eisheikh 2022, Stassen 2019] were also studied. Refer to Supplementary Information Section 2 for a more complete list of literature reviews.

However, increasingly complex challenges faced in urban areas require integrated analyses to identify potential gaps and overlaps in multiple services when they operate in tandem, aiming to maximize efficiency and build resilience. Understanding the impact of integrated ESS is essential for comprehending the resilience of emergency services and the preparedness of communities to handle complex emergencies that require multi-agency responses [Green 2017]. Studies such as [Zimmerman 2023, Kim 2021, Jeon 2018, Green 2017] have analyzed access to multiple emergency services. In the face of High-Impact, Low-Probability (HILP) events, such as Hurricane Sandy in 2012 [Aerts 2013], communities with high ESS accessibility rates and robust coverage of multiple services can recover from disasters more quickly, adding to the overall resilience of the system [Alexander 2015].

Fast response times depend on the physical accessibility of these services to the populace based on their geospatial locations and corresponding transportation networks. By considering the spatial context within which these emergency responses take place, the question of accessibility to emergency services can be framed as a transportation network connectivity problem [Novak 2014, Green 2017, Sohn 2006, Erath 2009].

This paper presents a scalable Geographic Information System (GIS) and network analysis-based approach to determine the accessibility of multiple emergency service sectors and to identify areas lacking emergency service access. Previous works disregarded the multi-component nature of ESS and treated it as a singular system, thereby lacking a holistic analysis. To address this gap, our study focused on evaluating the accessibility of three ESS components defined by CISA (Fire, EMS, Police), along with an additional service, Emergency Department (Hospital). We analyzed the site locations of these four sectors to understand not only the key metrics of each individual service, but also to holistically assess potential overlaps and gaps in services resulting from varying degrees of accessibility.

Furthermore, our aim is to determine the population residing in 'emergency service deserts', where multiple emergency services are inaccessible, rather than solely focusing on the size of the desert area. We utilized census data to investigate underlying causes for vulnerabilities and to suggest possible solutions to enhance overall accessibility. To the best of our knowledge, previous studies have not comprehensively examined more than three components of the ESS, and much of the literature has focused solely on identifying vulnerable regions without investigating the underlying causes of such vulnerabilities or proposing possible solutions to mitigate them.

To demonstrate this methodology, New York City was chosen due to its unique attributes: a highly diverse demographic landscape and a wide range of urban forms. Additionally, the city boasts one of the world's largest and most comprehensive ESS. Our research augments existing literature [e.g., Green 2017, Jeon 2018, Zimmerman 2023] by introducing a rapid assessment methodology, adaptable to different urban settings through the utilization of nationwide data sets, and open-source coding frameworks. This method further refines its analysis by downscaling census data to individual street intersections, enabling a highly granular assessment of service accessibility.

# 2. Methods

## 2.1 Data

This analysis relied on several essential datasets: (1) city and borough boundary dataset, (2) emergency services locations dataset, (3) road network dataset, and (4) census dataset.

### 2.1.1 City and Borough Boundaries

The city boundary of NYC, which defines the geographical domain of this study, was obtained from [CDC 2023]. The borough boundaries were from NYC OpenData [NYC 2023a]. Minor adjustments were made to the borough boundaries, such as the removal of small islands and the smoothening of piers and docks, to enhance visual clarity. Each borough is represented by a designated one-letter abbreviation: 'X' for Bronx, 'K' for Brooklyn (Kings County), 'M' for Manhattan, 'Q' for Queens, and 'R' for Staten Island (Richmond County). Additionally, a buffer zone of 7.5 km was added to avoid boundary effects and facilitate the analysis of regions where the closest ESS sites may lie just outside the NYC border. The boundaries of boroughs, city, and buffer zone are shown in Fig. 1A.

### 2.1.2 Emergency Services Site Locations

The site locations for each ESS service type are depicted in Fig. 2, along with corresponding boroughs. Locations for the four service types (Fire, EMS, Police, and Hospital) were sourced from the Homeland Infrastructure

Foundation-Level Data (HIFLD) [HIFLD 2020, 2021, 2023a, 2023b], based on the structure definitions used by the U.S. Geological Survey [USGS 2023]. To ensure the analysis is relevant to the general public, the dataset underwent modifications such as excluding children's hospitals, correctional facilities, and other ESS sites focused on specific groups of people (refer to the SI Section 1 for details on each sector). The breakdown of the number of ESS sites used in this study by the service type and borough is provided in Table 1.

### 2.1.3 Road Network

Road network for the domain and buffer regions were pulled from the OpenStreetMap (OSM) API via the OSMnx package[1] [Boeing 2017] in Python[2]. The road network was represented as a symmetric multigraph, and it was used to determine accessibility of ESS site locations via travel times along road networks. The roads used in this study are shown in Fig. 1B and a detailed description of the road network is in Subsubsection 2.2.1.

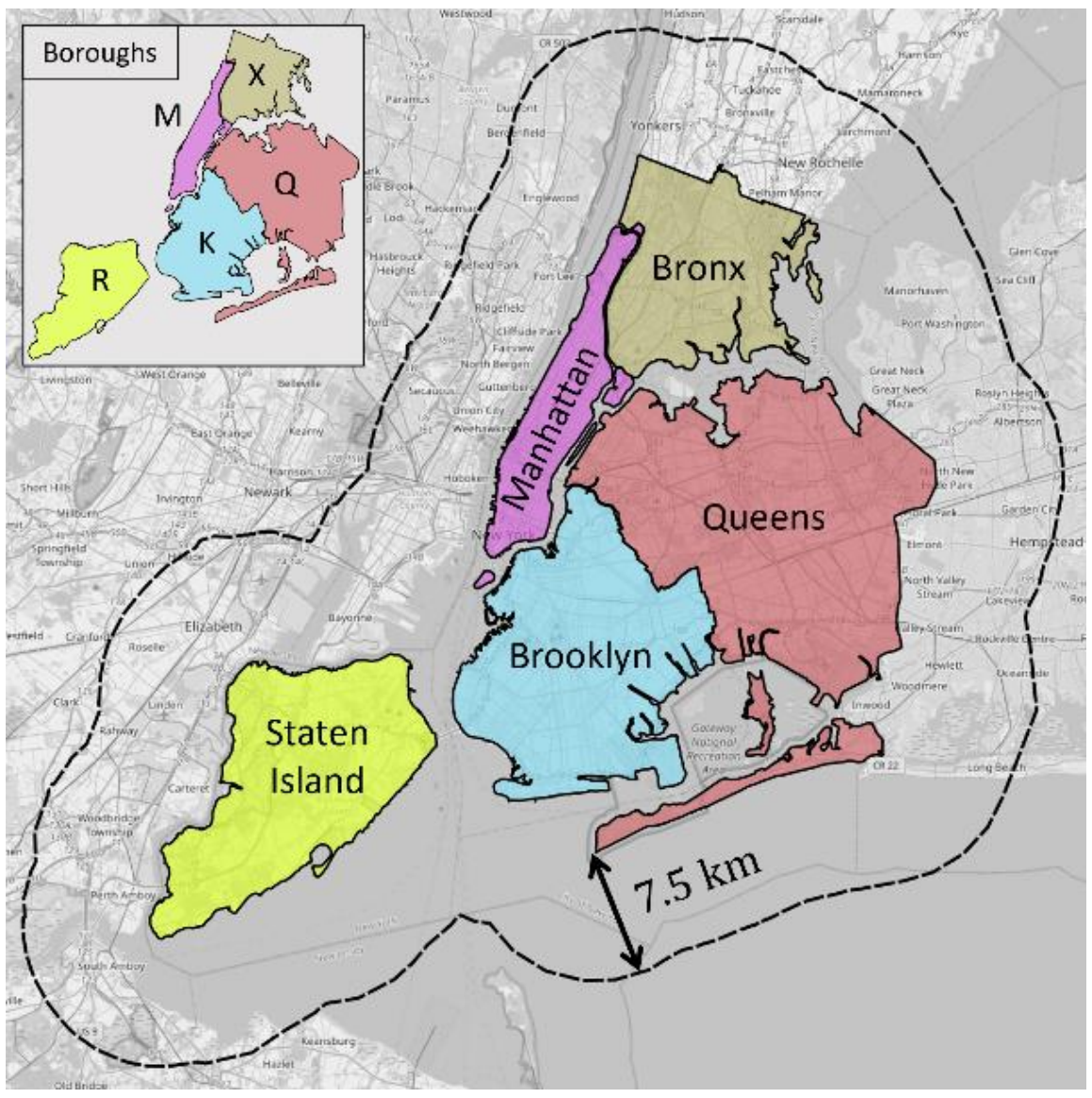

Fig. 1A: Representation of the geographical scope of the study, including the five boroughs of NYC. The inner solid black lines represent the domain boundary (NYC proper), while the outer dashed black line indicates the buffer boundary (7.5 km outward from the NYC boundary). The inset shows a simplified layout of the boroughs, each labeled with its one-letter abbreviation described in Subsubsection 2.1.1.

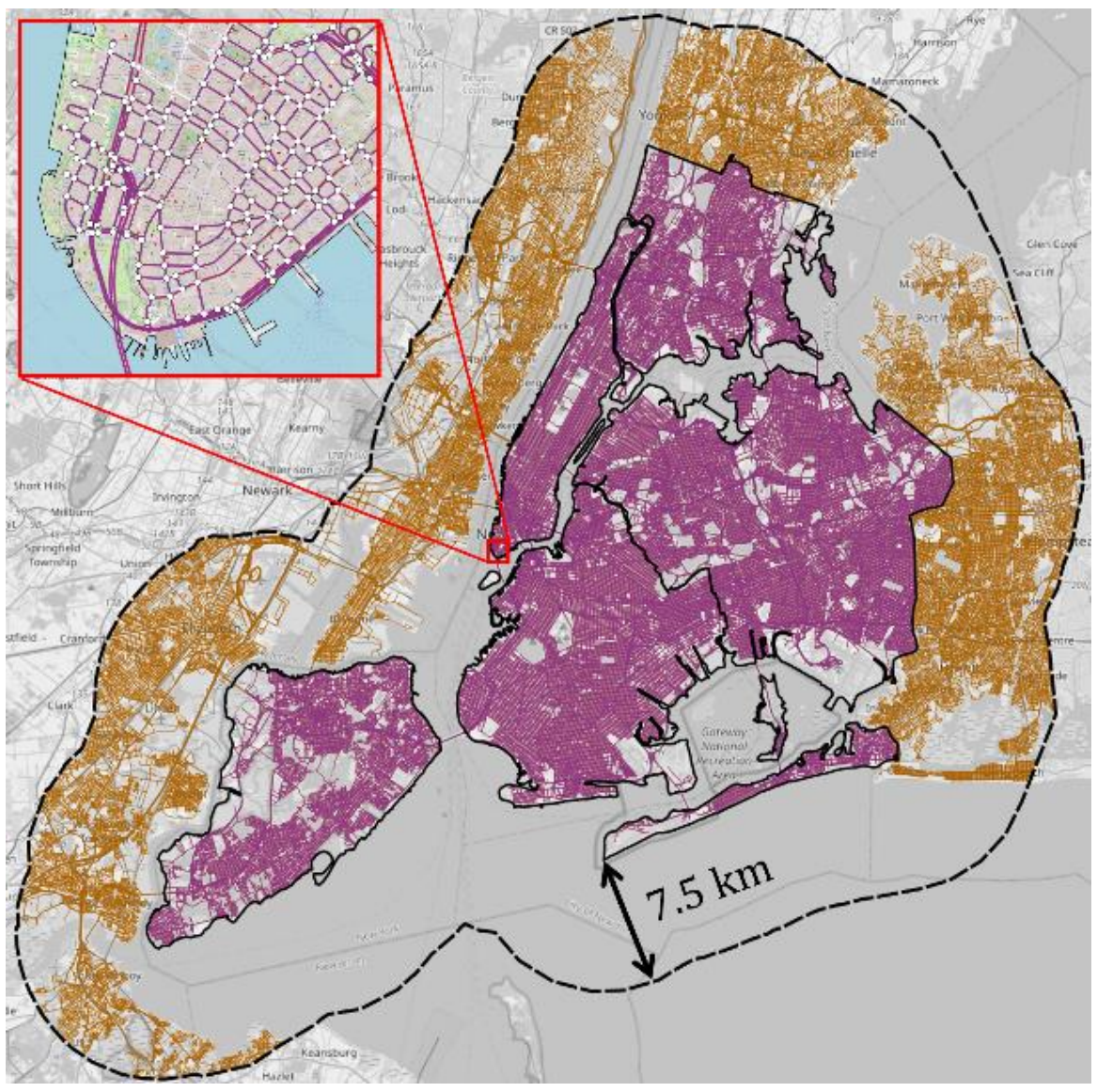

Fig. 1B: Representation of roads used in the analysis. Roads within the domain (NYC) boundary are in purple, while roads in the buffer zone are in brown. A close-up representation of the roads in lower Manhattan is enclosed within a red box, with white circles representing road network nodes (refer to Subsubsection 2.2.1).

---

1 Ver 1.2.2

2 Python 3.8.13

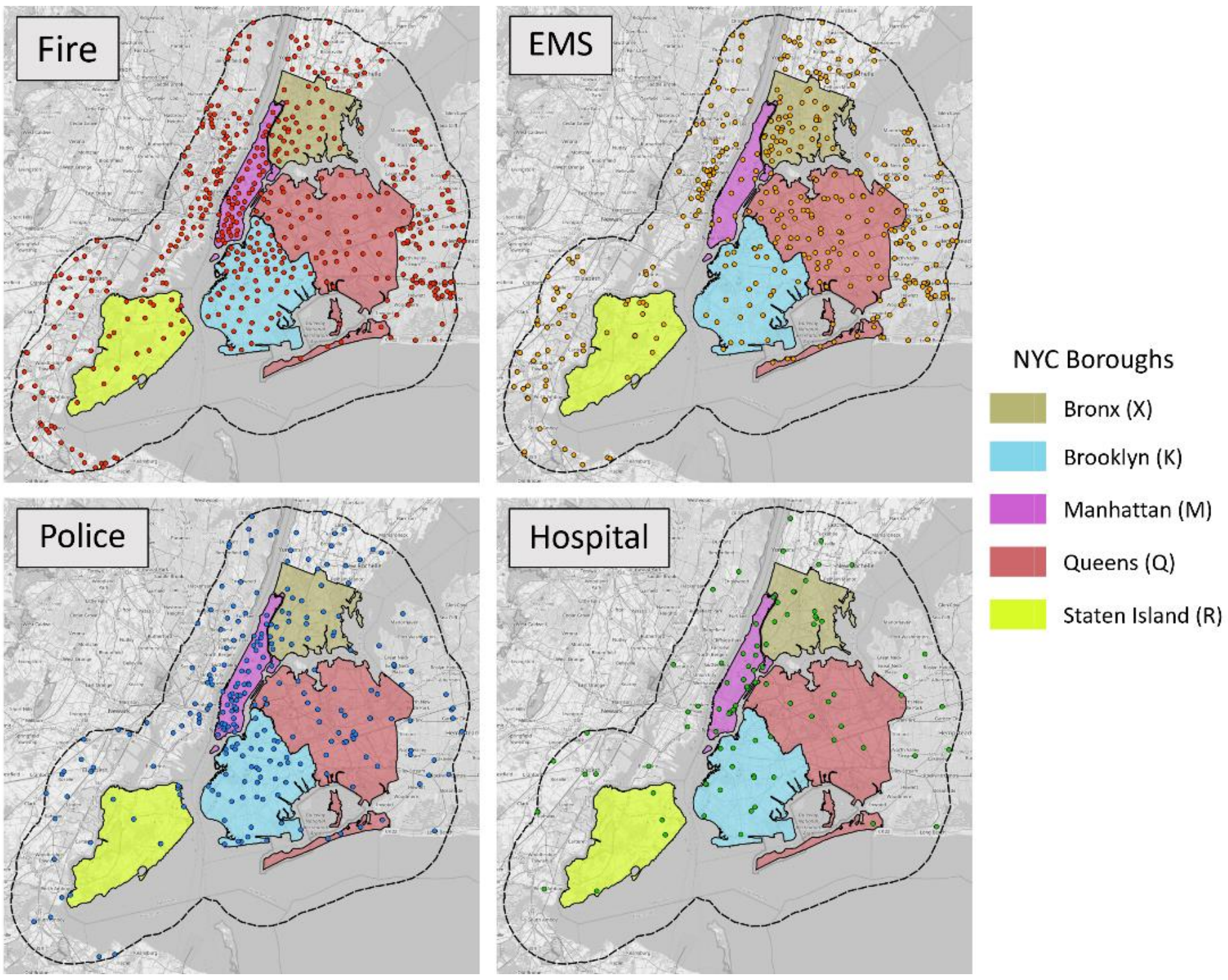

Fig. 2: Emergency Services Sectors Site Locations

Table 1: Number of ESS Sites for Each Service Type

| Emergency Service Sector | ESS Sites Represent | Number of Sites | | Number of Sites by Borough | | | | |
|---|---|---|---|---|---|---|---|---|
| | | Total | In NYC | Bronx | Brooklyn | Manhattan | Queens | Staten Island |
| Fire | Fire Stations | 481 | 222 | 34 | 66 | 49 | 51 | 22 |
| Emergency Medical Services | EMS Stations | 412 | 184 | 49 | 27 | 11 | 82 | 15 |
| Police | Police Stations | 257 | 166 | 24 | 47 | 47 | 39 | 9 |
| Hospital | Emergency Room | 91 | 64 | 12 | 15 | 23 | 10 | 4 |

### 2.1.4 NYC Census Information

Population data was used in this study to add more context to the question of ESS accessibility. Census data, including population per census tract, from the ACS 5-year averaged from 2016-2020 was accessed through the Census Bureau's API [USCB 2022b, 2023a]. Within the domain boundary, there is a population of 8,357,775 living within 2,353 census tracts. The spatial distribution of the populations and population densities of census tracts are shown in Fig. 3A and Fig. 3B, respectively. These values were aggregated to the borough level and are summarized in Table 2. It is important to note that population densities were calculated using the livable land area. This was calculated by excluding park [NYC 2023c] and water areas [USCB 2023b] from the original census tract areas. The calculated livable areas for each borough, along with their proportions, are presented in Table 2.

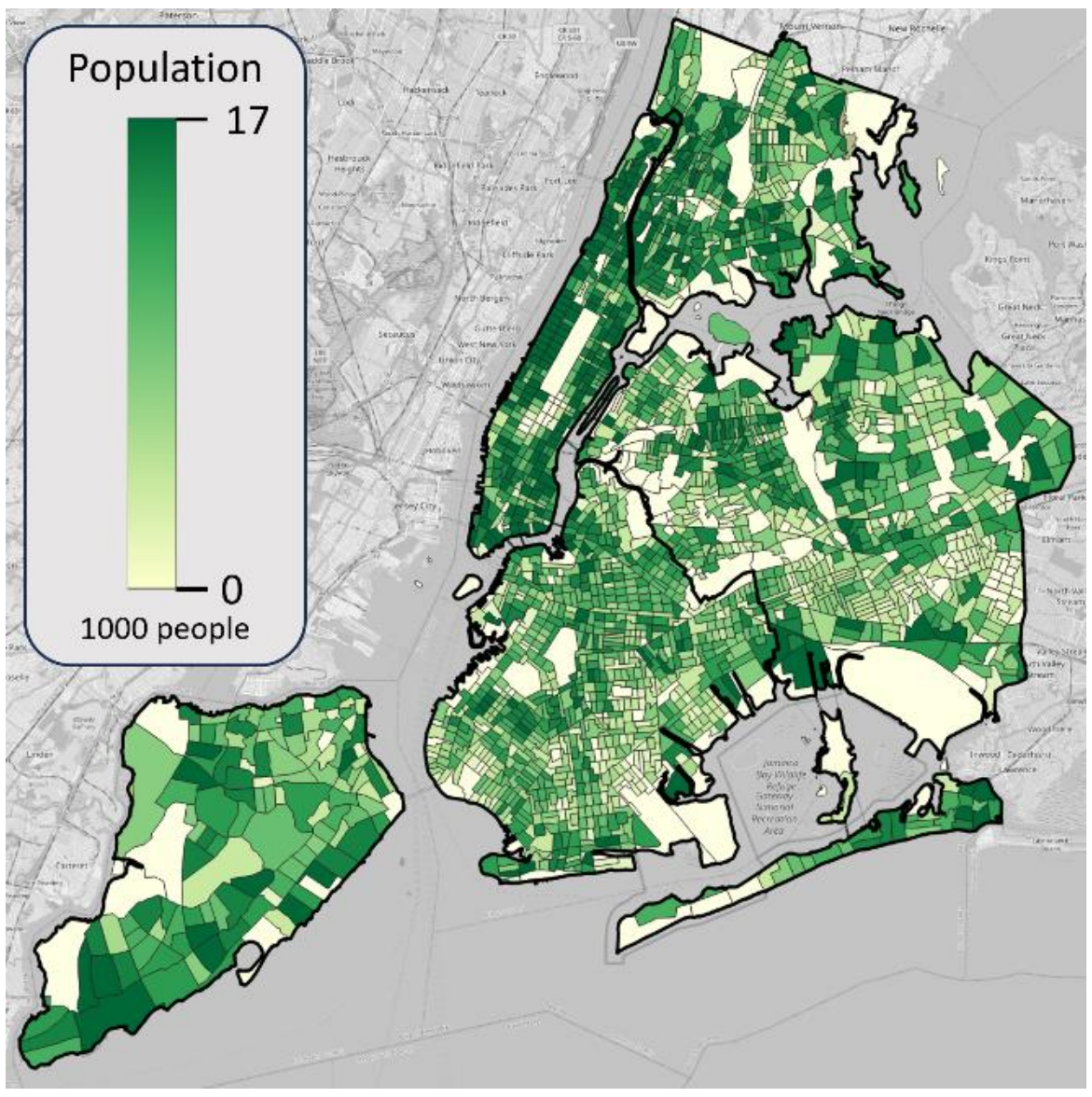

Fig. 3A: Population distribution of NYC census tracts.

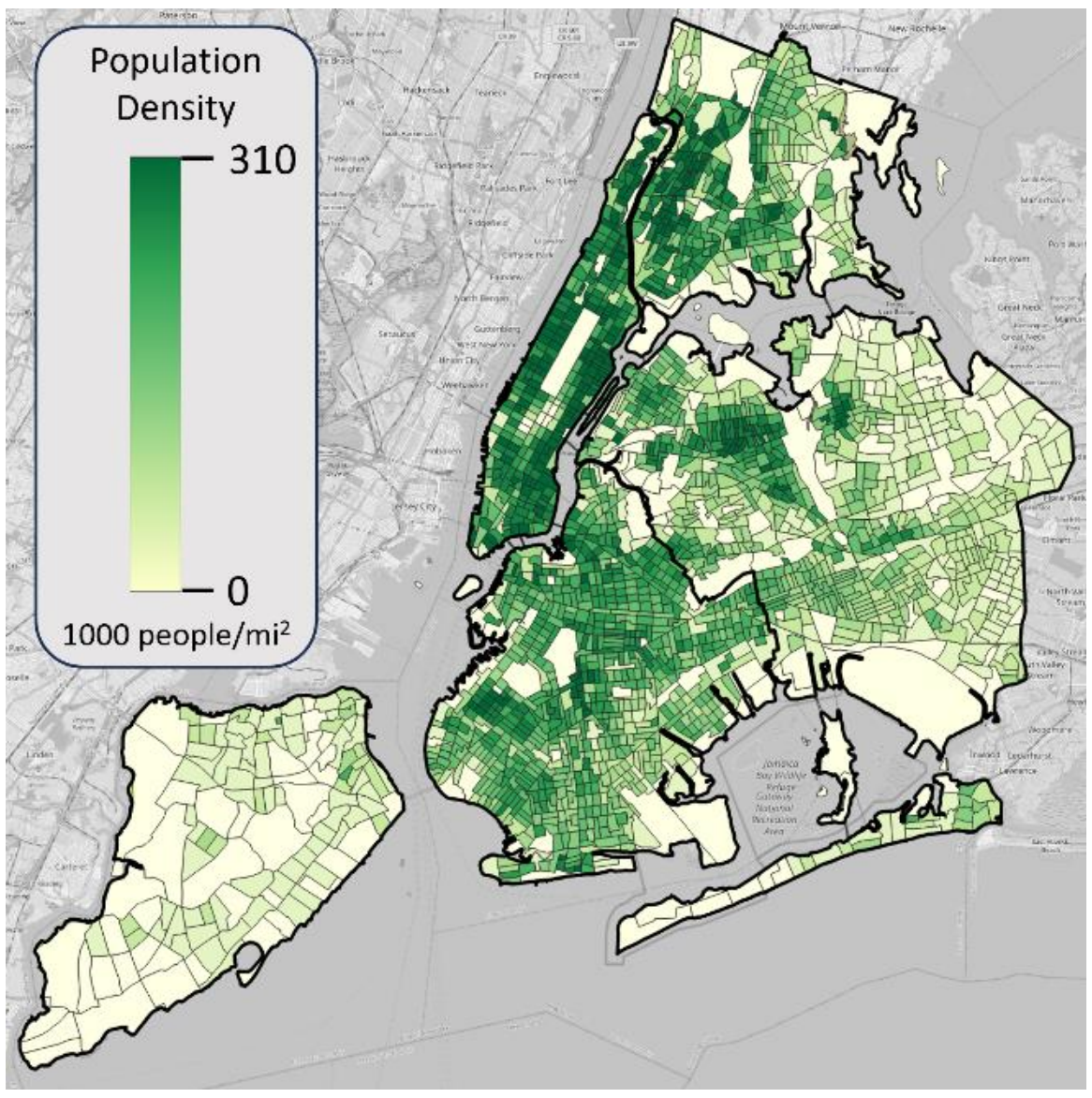

Fig. 3B: Population density distribution in livable lands of NYC census tracts.

Table 2: NYC Livable Land Area and Population Data

| | Bronx | Brooklyn | Manhattan | Queens | Staten Island | NYC |
|---|---|---|---|---|---|---|
| Livable Land Area [$mi^2$] | 32.32 (12.51%) | 62.32 (24.11%) | 18.46 (7.14%) | 97.72 (37.82%) | 47.60 (18.42%) | 258.42 |
| Population [million people] | 1.42 (16.97%) | 2.57 (30.80%) | 1.62 (19.38%) | 2.27 (27.17%) | 0.47 (5.67%) | 8.36 |
| Population Density [1000 people / $mi^2$] | 43.89 | 41.31 | 87.75 | 23.24 | 9.96 | 32.34 |

## 2.2 ESS Accessibility Model

Using the data gathered in Subsection 2.1, an ESS accessibility model was built for each service type. The process of building each model consisted of the following steps: (1) retrieving road network information, (2) identifying ESS site locations, (3) assigning population to regions, (4) calculating first responders travel times, and (5) determining accessibility and vulnerability of regions. The finalized ESS model includes information on population, population density, and travel time to the nearest ESS site for every region in the city.

### 2.2.1 Road Network Representation

The road network $G = (V, E)$ was represented by a set of nodes $V$ that are road intersections or cul-de-sacs, and a set of edges $E$ that are road segments between the nodes. The underlying assumption for the road network in this study is that people can only get on and off the road network at nodes (road intersections) and can only travel along the edges (road segments). Each node was assigned geographical coordinates, and each edge was a geographical line with attributes including a well-defined road length and speed limit. The nodes and edges along with their geographical locations were identified by OSM [OSM 2017]. For a visual representation of the road network, refer to Fig. 1B.

Road attributes such as road length, speed limit, and travel time were assigned by OSMnx. In many cases, speed limits were labeled by OSM, but missing speed limits were assigned as an average speed limit along the entire road by OSMnx [Boeing 2017]. Once the road lengths and speed limits were assigned, travel time for each edge was calculated by dividing the road length by the speed limit. Any disconnected road segments or isolated nodes, which may have arisen while clipping the roads with the buffer boundary, were removed to guarantee network connectivity. Also, edges were symmetrized by converting all one-way roads into bidirectional roads with the same speed limit because this study assumes that emergency vehicles can traverse through any road even if it is one-way for other drivers.

The resulting road network $G = (V, E)$ is a geographical connected symmetric graph. The final road network $G$ consisted of $|V| = 102{,}910$ nodes and $|E| = 162{,}072$ edges. The special subset of nodes $V_{NYC} \subset V$ within the NYC boundary was the focus of our study where $|V_{NYC}| = 55{,}329$ nodes.

### 2.2.2 ESS Site Nodes Identification

Using the ESS site locations data in Subsection 2.1, each ESS site was mapped to the geographically closest node $s \in V$ of the road network $G$ using Euclidean distance. The collection of these nodes, $S \subset V$, represents all the nodes that are closest to ESS sites. For example, $S_{Fire}$ represents all the fire stations and $S_{Police}$ represents all the police stations in $G$. For the geospatial distributions of $S_{Fire}$, $S_{EMS}$, $S_{Police}$, and $S_{Hospital}$, refer to Fig. 2. The road network $G$ and ESS site locations $S$ allow modeling a trip of first responders as a path between two nodes in the road network, where one of the origin or the destination must be from $S$.

Note that each ESS service type has its own accessibility model. Even though the underlying roads are unchanged, the difference in geospatial distribution of ESS site locations among different services yields different ESS site nodes, and thus yields a different accessibility model. For example, even though the police and firefighters use the same roads, the response times to the same location varies since the police stations and fire stations are distributed differently (refer to Fig. 5).

### 2.2.3 Travel Time Assignment

The travel time for a node to a specific emergency service was defined by the travel time from it to the closest ESS site. Specifically, given an ESS model $(G, S)$, the $|V_{NYC}| \times |S|$ travel time matrix $\mathbf{T}$ between the NYC nodes and all ESS nodes was calculated. The element $\mathrm{T}_{ij}$ of $\mathbf{T}$ represents the travel time needed to complete a path starting from a node $v_i \in V_{NYC}$ and ending at an ESS node $s_j \in S$. The path between $v_i$ to $s_j$ was calculated using the Dijkstra's algorithm minimizing the total sum of edge travel times. Since $G$ was connected by construction, every element of $\mathbf{T}$ was a finite non-negative value. Once the calculation of $\mathbf{T}$ was complete, taking the row-wise minimum on $\mathbf{T}$ yielded the fastest travel time vector

$$\mathbf{t} = \left(t_1, \dots, t_i, \dots t_{|V_{NYC}|}\right) := \min_j \mathrm{T}_{ij}.$$

Here, $t_i$ represented the travel time required to reach the closest ESS location from the node $v_i \in V_{NYC}$.

While $\mathbf{T}$ represents only travel times ending at ESS nodes, another $|S| \times |V_{NYC}|$ travel time matrix $\mathbf{T}'$ to represent the other direction was not needed due to the symmetric nature of the base graph. In reality, the travel time from a node $v$ to a node $u$ can be different from the travel time from $u$ to $v$. However, since $G$ was constructed to be symmetric with the same speed limits and road lengths in both directions, $\mathbf{T} = (\mathbf{T}')^T$ is always satisfied.

For this study, a region's accessibility to an ESS was purely determined based on the fastest travel time from the region to any of the ESS site. From now on, unless specified otherwise, the term 'travel time' refers to the fastest travel time to ESS sites and the 'closest site' is determined by the travel time, not by the distance.

### 2.2.4 Population Assignment

A pure distance based or travel time-based accessibility of ESS may underrepresent people living in a densely populated region. To overcome this issue, it is necessary to consider the population density of regions and determine ESS accessibility based on the total number of residents living in inaccessible areas, rather than just the geographical area.

The set of nodes $V_{NYC}$ represents locations where people access and exit the road network. Under the assumption that areas with higher population density generate more travel, each node was weighted by the population density. To achieve this, NYC was partitioned into Voronoi regions using the set of road network nodes as seeds. Each node $v \in V_{NYC}$ was associated with a Voronoi region $R(v)$, consisting of all points in NYC closer to $v$ than to any other nodes in $V_{NYC}$ [Burrough 2015]. Because of this, $R(v)$ represents the area of influence of node $v$ in NYC, with the number of residents in $R(v)$ indicating the population using $v$ to access and exit the road network.

A census tract $c \in C$ represents a region where demographic information like population $P(c)$ is known for the purpose of taking a census [USCB 2022a]. Instead of using a simple $\rho(c) = P(c)/Area(c)$ as a population density, non-residential areas like parks and water areas from census tracts were removed to calculate the livable land area $L(c)$. The adjusted population density $\rho^*(c) \coloneqq P(c)/L(c) \geq \rho(c)$ represents a more accurate population density than $\rho(c)$ for each census tract $c$.

The areal influence $a(v, c) = |R(v) \cap c|$ of each node on each census tract was calculated by geospatially intersecting the Voronoi region $R(v)$ and the census tract $c$ [Horner 2010]. This step is necessary because not all Voronoi regions are completely contained in a single census tract. Often census tract boundaries are defined by road segments, which forces some road network nodes to be on the census tract boundaries, causing associated Voronoi regions to overlap with multiple census tracts. To appropriately assign population to such nodes, all the census tracts which the Voronoi region overlaps with need to be weighted proportionally based on the overlapping area. Finally, the total population to each node is defined by

$$P(v) \coloneqq \sum_{c \in C} a(v, c) \cdot \rho^*(c) = \sum_{c \in C} |R(v) \cap c| \cdot \frac{P(c)}{L(c)}$$

where $P(v)$ is the total population associated with node $v$, $a(v, c)$ is the overlapping area of the Voronoi region associated with the node $v$ and census tract $c$, and $\rho^*(c)$ is the population density of livable land area in census tract $c$. With this method, the entire livable area and the total residents of NYC were partitioned into $|V_{NYC}| = 55{,}329$ nodes.

### 2.2.5 Accessibility and Vulnerability Determination

The accessibility of a region to each ESS component depends on whether there is an ESS site that first responders can reach within a benchmark time $\tau$. For each node $v_i \in V_{NYC}$, the node has access to a specific ESS component if the calculated travel time satisfies $t_i \leq \tau$. In general, an appropriate choice of $\tau$ depends on the geographical scope of the study and the ESS component concerned, but the choice of $\tau = 4$ minutes in this study aligns with the goal set by the NFPA Standard 1710 [NFPA 2020] to benchmark emergency response times. For each ESS service type, this determination separates the region of study into accessible and inaccessible regions (Fig. 5). Once accessibility is determined for each ESS, aggregating them illuminates vulnerable regions since a single emergency service alone is not sufficient to provide a safe community. Safeguarding human life and property can be more effectively done if people have access to multiple ESSs. The more ESS access a region has, the less vulnerable the region is. Finally, once the accessibility and vulnerability of regions are determined, the number of people impacted by inaccessibility and vulnerability can be calculated by summing up the assigned population in those identified inaccessible and vulnerable region.

# 3 Results

## 3.1 Travel Times

Fig. 4 displays the probability density function of travel times for each ESS service type. All density functions are unimodal but lack symmetry due to the presence of long tails, which are associated with regions with significantly higher travel times for emergency services to access. Because of these long tails, the median is a more appropriate representation of travel times than the mean value. The order in which median travel times increase between services (Fire < EMS < Police < Hospital) is inverse to the order of how number of unique ESS sites increases (Hospital < Police < EMS < Fire), when referring to the median travel time values in the inset table of Fig. 4 and the number of ESS sites in Table 1.

The accessibility maps in Fig. 5 present distribution of travel times to the nearest ESS site. The blue regions indicate areas where at least one ESS site location can be reached within $\tau = 4$ minutes, while the red regions signify areas where no ESS locations can be reached within the same time frame. While the majority of NYC regions have access to Fire services, a significant portion of the city lacks access to Hospitals within a 4-minute travel time.

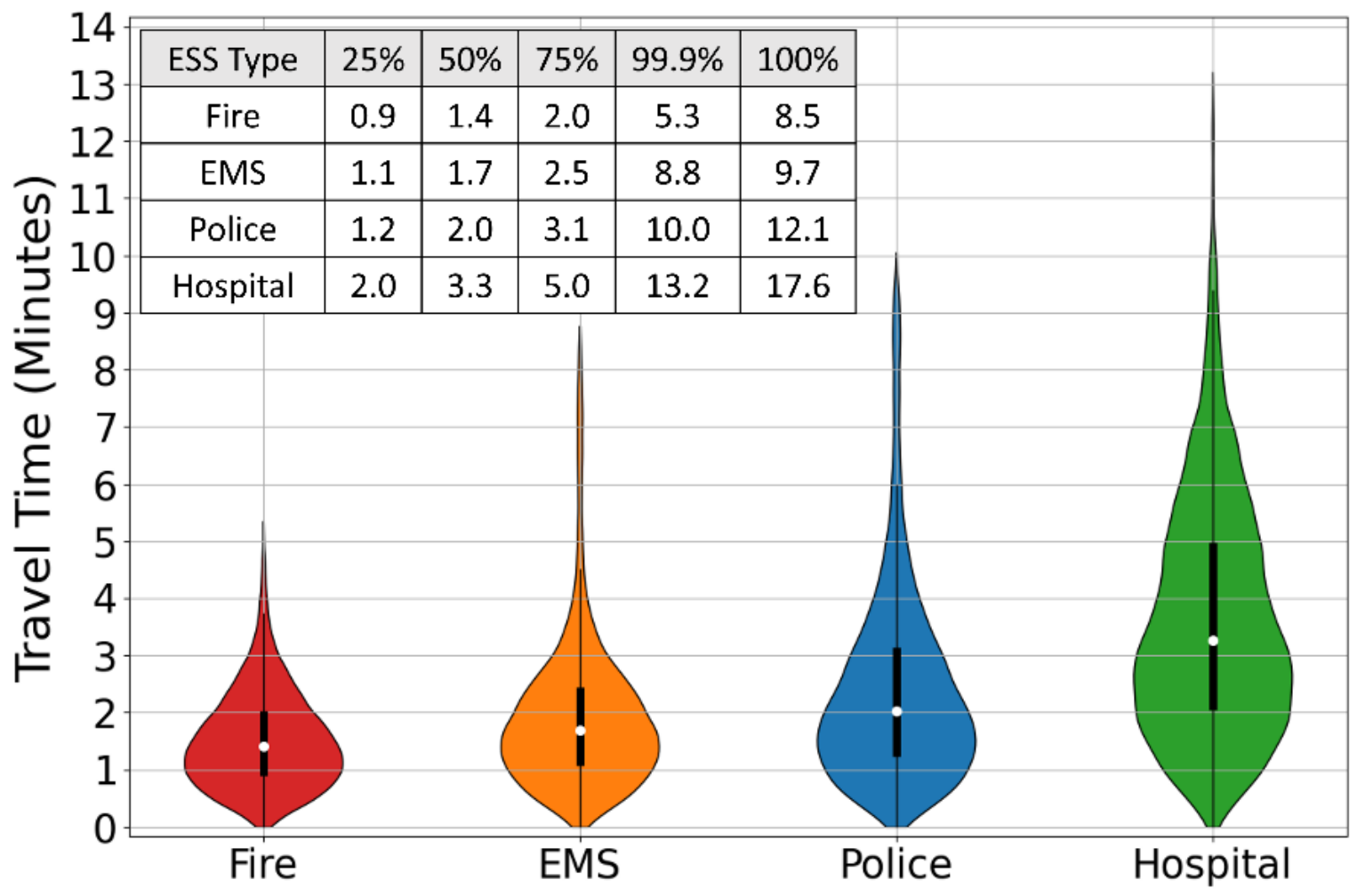

| ESS Type | 25% | 50% | 75% | 99.9% | 100% |
|---|---|---|---|---|---|
| Fire | 0.9 | 1.4 | 2.0 | 5.3 | 8.5 |
| EMS | 1.1 | 1.7 | 2.5 | 8.8 | 9.7 |
| Police | 1.2 | 2.0 | 3.1 | 10.0 | 12.1 |
| Hospital | 2.0 | 3.3 | 5.0 | 13.2 | 17.6 |

Fig. 4: The violin plots represent the probability density function of 99.9% of travel times for each ESS. The longest 0.1% of travel times were excluded from display to avoid extremely long tails (refer to the 99.9% and 100% values in the inset table). The median and interquartile range are shown as white dots and black solid lines inside the violins. The values of 25%, 50% (median), 75%, 99.9%, and 100% (maximum) travel times in minutes are detailed in the inset table.

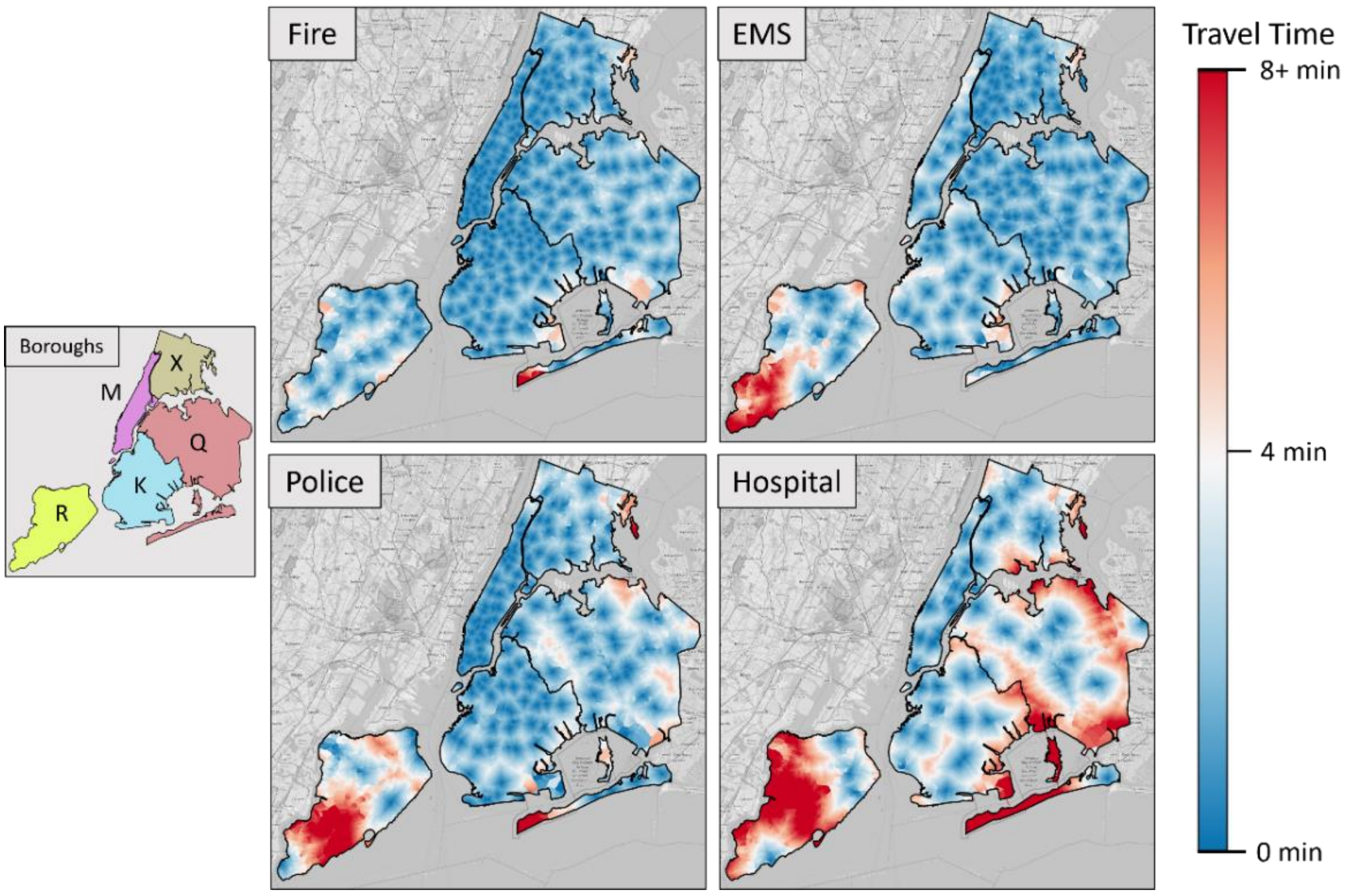

Fig. 5: Accessibility maps representing travel times to the nearest ESS site location indicated by color. Two different color spectra (blue and red) were used to indicate the intensity of accessibility (blue) and inaccessibility (red), with the neutral color (white) to represent the benchmark time $\tau = 4$ minutes. Any regions with the travel time greater than or equal to 8 minutes were represented by the same color to preserve the symmetry of the color bar.

## 3.2 Service Availability and Vulnerable Regions

To observe regions that are commonly lacking emergency services, the accessible (blue) regions and inaccessible (red) regions in Fig. 5 were overlayed to produce the number of available emergency services for each region. In Fig. 6, most of NYC has access to 3 or 4 emergency services and is well-served. However, Staten Island (represented by 'R' in the Boroughs inset map) and Queens (represented by 'Q' in the Boroughs inset map) have the majority of underserved regions with 0, 1, or 2 emergency services. These underserved regions are highlighted in Fig. 7.

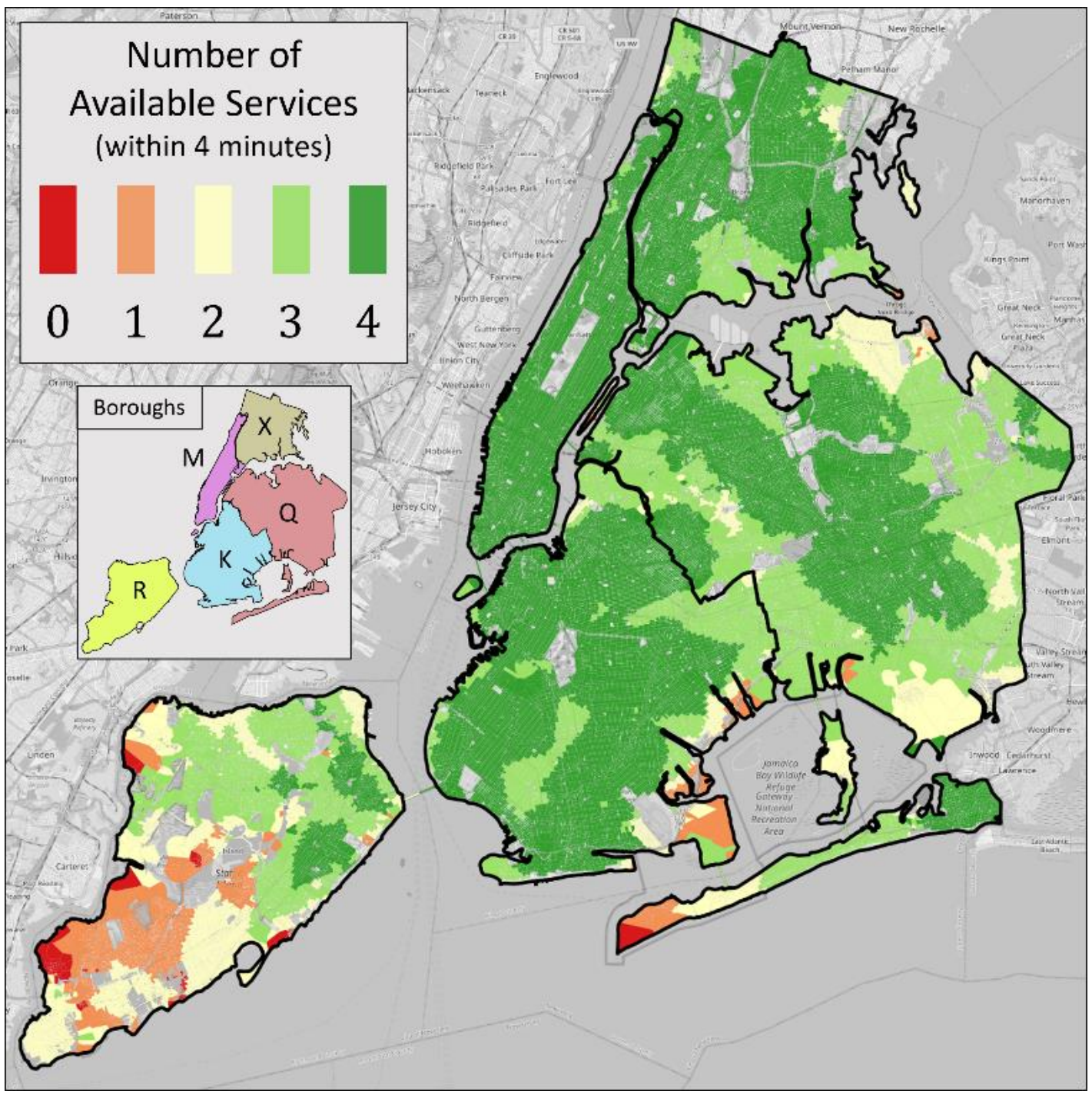

Fig. 6: A vulnerability map indicating well-served regions in green (3 or 4 available services) and underserved regions in red, orange, and yellow (0, 1, or 2 available services). Parks and water areas were removed for more accurate analysis.

The $\tau = 4$ minute travel time benchmark used to determine the accessibility of emergency services was a matter of definition. To identify vulnerable regions more clearly, a second vulnerability map (Fig. 7) was created with a modified travel time cutoff $\tau' = \tau + 1$ minute to highlight only the vulnerable regions. With this additional minute of travel time, regions that remain inaccessible are more confidently labeled as such. Fig. 7 illustrates five geographically distinct vulnerable regions, aligning with the vulnerable regions in Fig. 6 as expected. Fig. 7A-E provide close-ups of each vulnerable region, accompanied by their respective names. The five vulnerable regions are Roosevelt Island in East River, Northern Queens, City Island in the east of Bronx, Staten Island, and Jamaica Bay in the south of Queens.

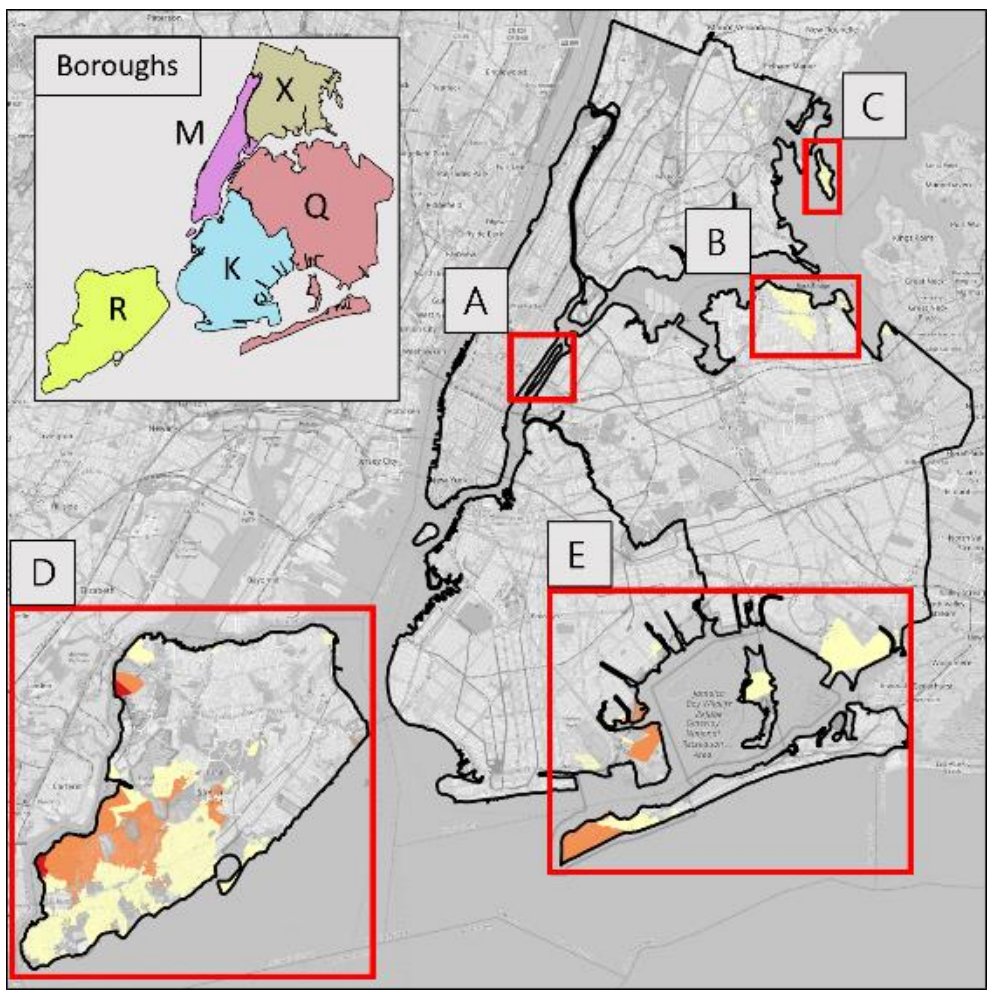

Fig. 7: Vulnerable regions in NYC based on the 5 minutes benchmark time.

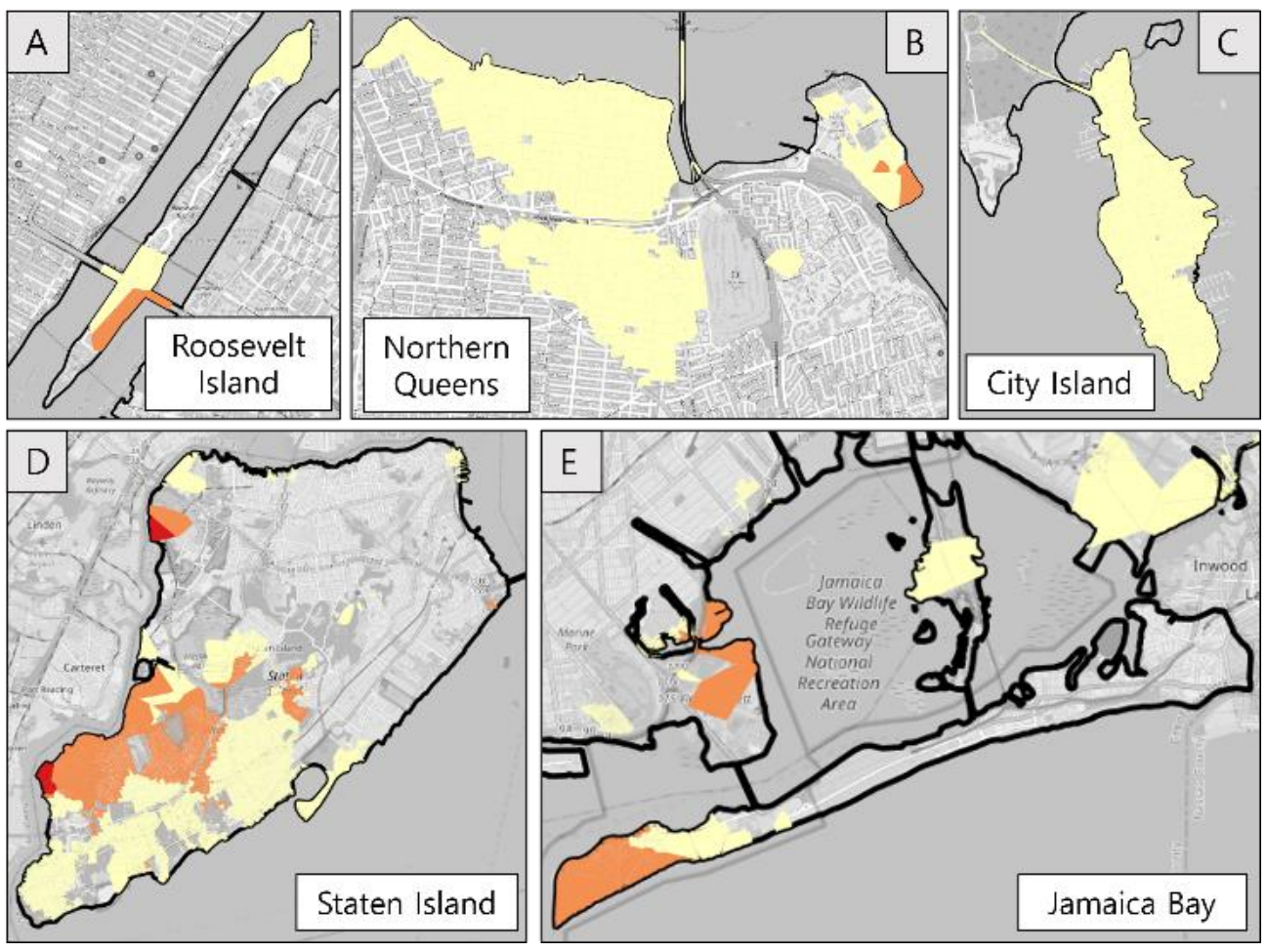

Fig. 7A-E: Close up of the five vulnerable regions with their region names. (not to scale)

Using the population density assigned in Subsubsection 2.2.4, the number of residents with varying degrees of ESS accessibility is summarized in Table 3. The majority of NYC residents (76%) have access to all emergency services, and the absolute majority (95%) are well-served. On the other hand, a significantly small number of people (5%) are underserved, and almost none of the residents (0.05%) have no services. This indicates that the current ESS effectively serves NYC residents.

However, disaggregating the result by boroughs reveals inhomogeneity in emergency service accessibility. Referring to Table 3, well-served residents constitute the absolute majority of residents (99%) in Manhattan, which is the densest borough. This trend is also observed in the Bronx and Brooklyn where 98% of residents are well-served, and in Queens where 95% of residents are well-served. In contrast, Staten Island, the sparsest borough, has only 53% of its residents considered well-served.

Table 3: Number of People with Access to Different Number of Emergency Services

| Vulnerability | Available Services | Population | Population Proportion | Breakdown by Borough | | | | |
|---|---|---|---|---|---|---|---|---|
| | | | | Bronx | Brooklyn | Manhattan | Queens | SI |
| Not Vulnerable (Well-served) | 4 | 6,349,282 | 75.97% | 14.45% | 24.70% | 19.02% | 16.96% | 0.82% |
| | 3 | 1,587,219 | 18.99% | 2.20% | 5.45% | 0.27% | 8.86% | 2.21% |
| Vulnerable (Underserved) | 2 | 319,659 | 3.82% | 0.31% | 0.55% | 0.07% | 1.30% | 1.59% |
| | 1 | 97,048 | 1.16% | 0.01% | 0.10% | 0.02% | 0.04% | 1.00% |
| | 0 | 4,567 | 0.05% | 0.00% | 0.00% | 0.01% | 0.00% | 0.05% |
| | Total | 8,357,775 | 100% | 16.97% | 30.80% | 19.39% | 27.16% | 5.67% |

## 3.3 Relationships Between Travel Time, ESS Site Density, and Population Density

This subsection presents pair-wise relationships between travel time $T$, ESS site density $\rho_{ESS}$, and population density $\rho_{pop}$. All data was collected for each region (5 boroughs + NYC as a whole) and for each ESS type (Fire, EMS, Police, and Hospital). When calculating the trend of data points, such as the curve of best fit or Pearson correlation, only the borough data were used. Since the borough data and the entire NYC data are not independent, only the 20 data points (5 boroughs $\times$ 4 ESS types) were used in trend calculations.

The $\rho_{ESS}$ used here is not simply the ratio of the number of ESS sites and the livable borough area. Instead, it is calculated as the ratio of the number of 'reachable' ESS sites to the livable borough area as defined below:

$$\rho_{ESS} = \frac{\#\, Reachable\ ESS\ Sites}{Livable\ Area} \geq \frac{\#\, ESS\ Sites}{Livable\ Area}.$$

To mitigate boundary effects between boroughs, the number of reachable ESS sites were used, representing the number of unique ESS sites that each borough has access to, even those located just outside of the borough boundary.

### 3.3.1 Travel Time and ESS Site Density

Since all the boroughs have different land areas, the actual number of ESS sites cannot be compared directly between boroughs. Instead, the median travel times $T$ and ESS site density $\rho_{ESS}$ for each region are shown in a log-log scale in Fig 8. The diagonal black dashed line in Fig. 8 represents the power law $T \propto (\rho_{ESS})^{-\alpha}$ that best fits the data across all boroughs and all ESS types. The value of $\alpha = 0.59$ indicates that every time the ESS site density doubles, the median travel times to that ESS decreases by a factor of $2^{0.59} \approx 1.51$ in NYC. While it would be interesting to explore how the $\alpha$ values vary by boroughs, having only four data points per borough does not allow for a statistically significant fitting of a curve.

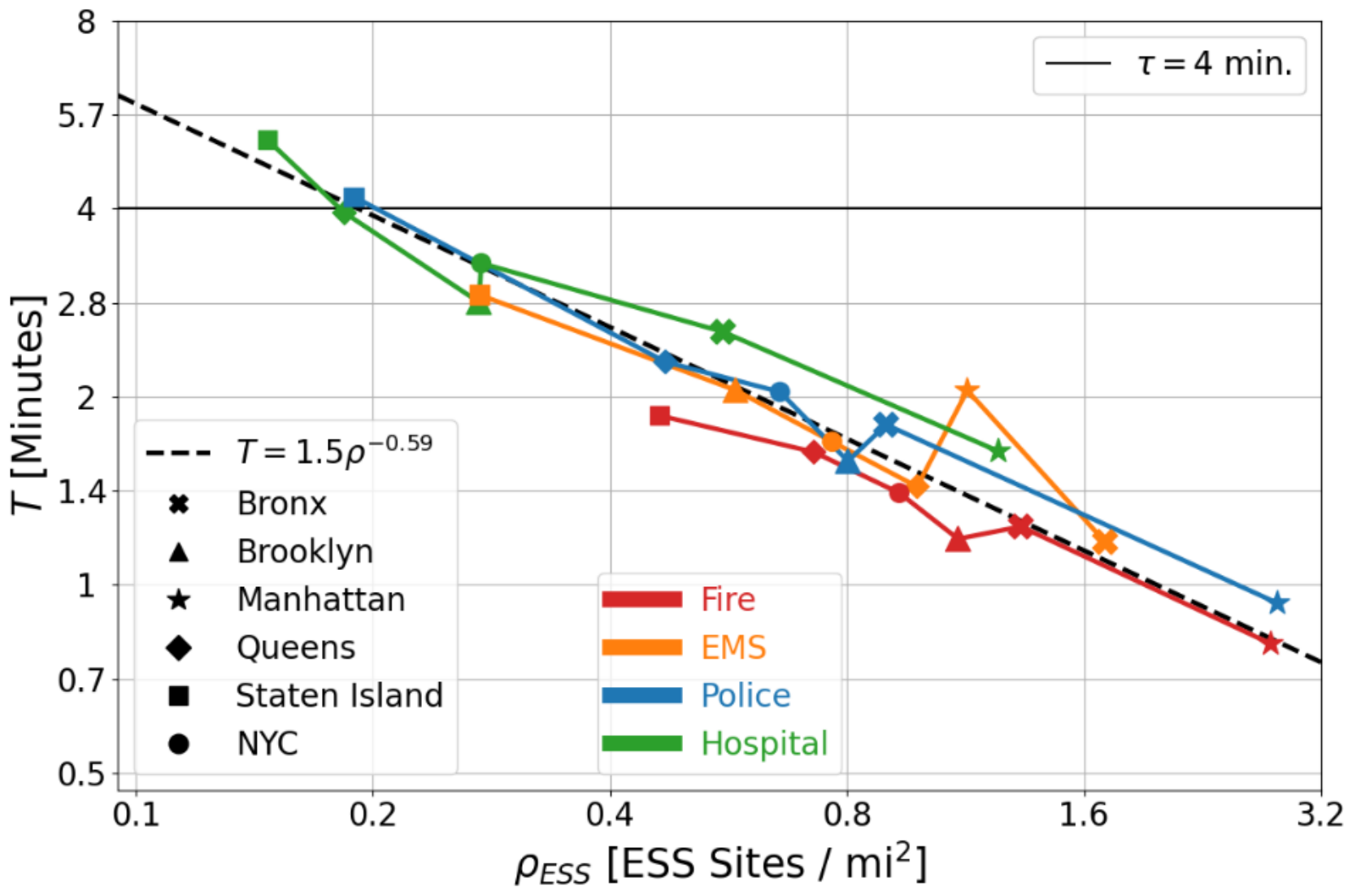

Fig. 8: The relationship between median travel times $T$ and ESS site density $\rho_{ESS}$ is illustrated in a log-log plot. Different symbols denote different geographical regions (refer to the lower left legend), while distinct colors signify different ESS types (refer to the lower center legend). The diagonal black dashed line represents the fitted power law $T = 1.5 \cdot (\rho_{ESS})^{-0.59}$ with $R^2 = 0.93$. The horizontal black solid line indicates the benchmark time $\tau = 4$ minutes (refer to the upper right legend).

### 3.3.2 ESS Site Density and Population Density

ESS sites density and population desnity are likely to be considered in the siting of ESS services. City managers are not likely to put ESS sites far away from the populations they serve. To correlate $T$ and $\rho_{pop}$, a relationship between $\rho_{ESS}$ and $\rho_{pop}$ (Fig. 9) is needed to convert the results from Fig. 8. Unlike the power law relationship found in Subsubection 3.3.1, a specific model could not be fit in this. However, there still is a positive Pearson correlation ($r = 0.78$) between $\rho_{ESS}$ and $\rho_{pop}$. Observing that EMS behaves differently than other ESS types, excluding EMS in Pearson correlation calculation yields a stronger correlation of $r' = 0.85$.

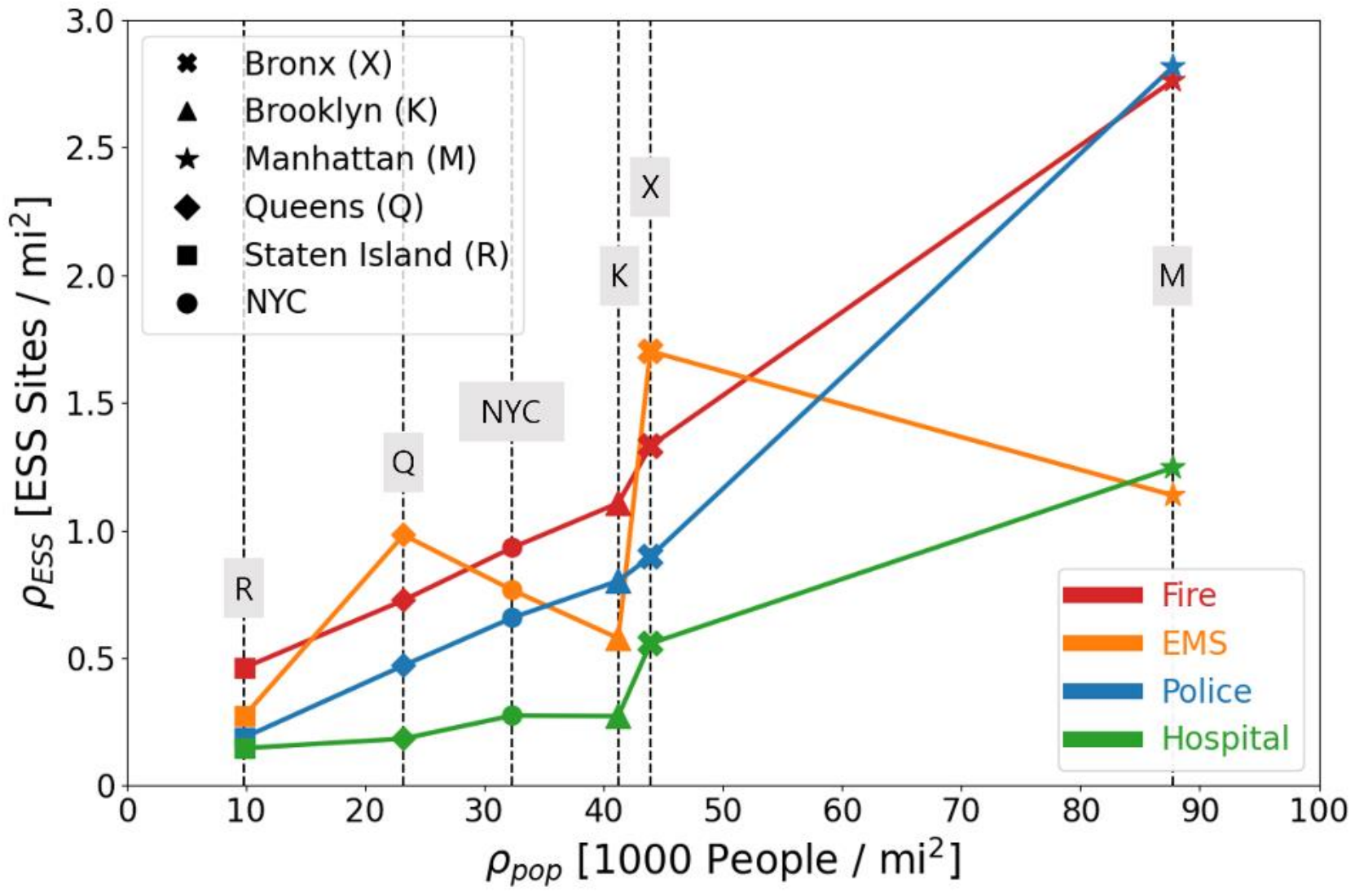

Fig. 9: The relationship between ESS site density $\rho_{ESS}$ and population density $\rho_{pop}$. Different symbols denote different geographical regions (refer to the upper left legend), while distinct colors signify different ESS types (refer to the lower right legend). Each vertical black dashed line represents data for a specific region along with their one-letter abbreviation for the boroughs. The Pearson correlation coefficient using data points from all four services is $r = 0.78$, and using data points from Fire, Police, and Hospital only, it is $r' = 0.85$. Data from NYC as a whole was excluded when calculating Pearson correlations.

### 3.3.3 Travel Time and Population Density

Combining the results from Subsubsections 3.3.1 and 3.3.2, the travel times $T$ and population density $\rho_{pop}$ are plotted in log-log scale in Fig. 10. Like Subsubsection 3.3.2, a specific model could not be fit to this due to a large variance, but negatively correlated trend is still shown between $T$ and $\rho_{pop}$ for each ESS type, except EMS. This suggests that population density of where one lives can impact one's accessibility of emergency services.

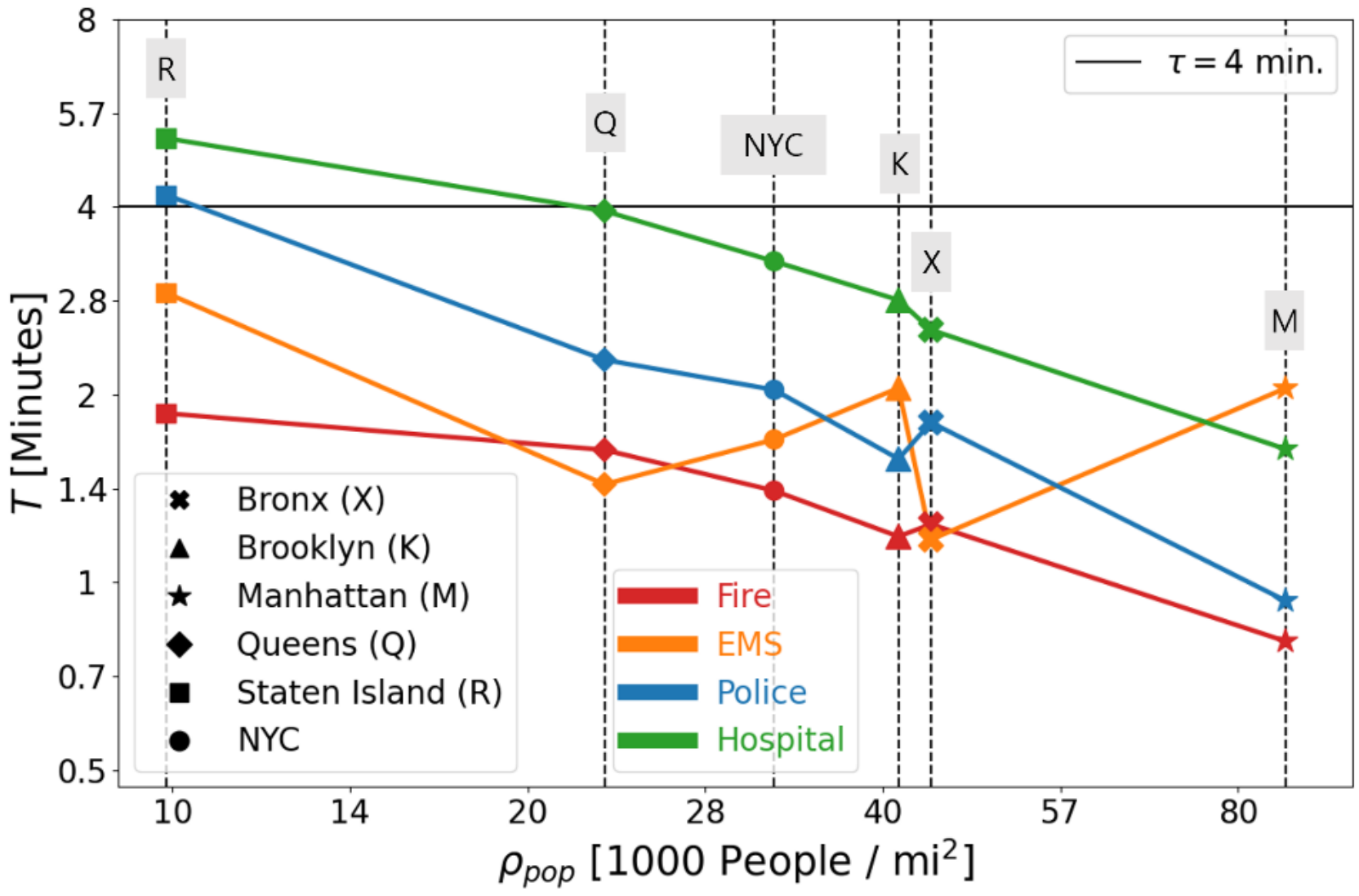

Fig. 10: The relationship between median travel times $T$ and population density $\rho_{pop}$ in a log-log plot. Each vertical black dashed line represents data for a specific region. The Pearson correlation coefficient for all four services is $r = -0.57$, and for Fire, Police, and Hospital only, it is $r' = -0.65$. Data from NYC as a whole was excluded when calculating Pearson correlations.

# 4 Discussions

Comparing the results from Section 3 with real-world metrics, the model reasonably mirrored actual travel times observed in the NYC 911 data [NYC 2023b] and the chosen benchmark was deemed feasible for NYC. Subsequently, an examination of vulnerable regions highlighted in Fig. 7 revealed that many of these areas exhibited geographical bottlenecks or low population density. An exploration of accessibility and population density suggested that enhancing overall accessibility in NYC could be efficiently achieved by investing in additional sites in less populated areas, contrary to common assumptions.

## 4.1 Model Comparison to Real World Metrics

### 4.1.1 Actual Travel Time

NYC 911 Reporting data is made public [NYC 2023b], and it is compared to the transportation network modeling done here. The End-to-End Detail Report provides the average and median timestamps for each segment of the entire 911 call from first pickup of the call to first arrival of responders. The time difference between “Agency Dispatch” and “Agency Arrival” timestamps was used to estimate the average and median travel times of the top priority emergency responses. The first three columns of Table 4 summarize the average and median travel times for the actual top priority calls in NYC in the period of Jan. 01, 2023, to Jul. 02, 2023 [NYC 2023b]. The top priority calls included life threatening medical emergencies, structural fires, and critical police calls. Since the actual 911 call data only included FDNY, EMS, and NYPD agencies, it is not clear how to compare the simulated Hospital travel data to the actual data. For this reason, Table 5 only includes the simulated Fire, EMS, and Police results.

Table 4: Simulated and Actual Travel Times in Minutes Compared

| NYC Agency | Actual Avg. T | Actual Med. T | ESS Type | Simulated Avg. T | Simulated Med. T | $\frac{\text{Avg Actual}}{\text{Avg Simulated}}$ | $\frac{\text{Med Actual}}{\text{Med Simulated}}$ |
|---|---|---|---|---|---|---|---|
| FDNY (Fire) | 3.33 | 3.33 | Fire | 1.54 | 1.40 | 2.16 | 2.38 |
| FDNY (Medical) | 6.08 | 5.97 | EMS | 1.96 | 1.70 | 3.10 | 3.58 |
| EMS | 7.18 | 6.75 | | | | 3.66 | 3.97 |
| NYPD | 4.53 | 3.42 | Police | 2.44 | 2.04 | 1.86 | 1.68 |

Table 5 shows that the simulated results underestimate actual travel times given by the NYC 911 Calls Reporting data [NYC 2023b]. The actual travel times are 1.68 times (Police) to 3.97 times (EMS) longer than the simulated results. The simplest way to adjust our model is to scale all speed limits of roads by a single factor to best fit the simulated values to the actual values. However, this is only reasonable if all our simulated values were off by the same percentage as the actual values. Since each ESS type underestimates the actual travel times by a different factor, it shows that modeling the modeling the true travel time for dispatches requires more than just a road network and speed limits. Instead, it depends on a variety of factors, such as traffic conditions and the existence of shoulders on roads.

Despite the underestimation of travel times, they remain within the same order of magnitudes. The analysis remains valuable as it identifies vulnerable regions even under the best-case scenario of NYC with no traffic. Future studies could explore incorporating congestion traffic models, such as [Ganin 2017], to bring modeled and observed values more closely. However, achieving a realistic traffic model would require additional datasets.

When considering traffic, the presence of shoulders can significantly reduce travel time in many cases as first responders often use shoulders to access crash sites on busy congested highways. Incorporating shoulder information into our model can be achieved with minimal adjustment, as OpenStreetMap already includes the ‘shoulder’ key to indicate the existence of breakdown lanes along with their width. However, utilizing shoulders to facilitate faster emergency vehicle response requires more detailed information beyond their mere presence and width.

In densely populated residential or commercial areas, such as downtown Manhattan, shoulders on roads are often used for street parking and by vendors. The presence of a single parked car or a dining table makes the shoulders ineffective for expediting emergency responses. Because of this, the efficacy of shoulders may be limited in densely populated areas, whereas they can prove effective in sparsely populated regions or on highways, particularly when considering congestion. To mitigate street parking issues, one may use the ‘parking’ tag on OSM to offset the effect of shoulders. However, this approach can be unreliable since street parking is heavily influenced by many temporary factors, such as constructions and events.

### 4.1.2 Benchmark Travel Time

Even though $\tau = 4$ minutes was chosen as the benchmark in determining accessibility based on the NFPA Standard 1710 (NFPA 2020) guideline, the feasibility for NYC was unclear. By varying $\tau$, it is possible to observe how accessibility to each ESS type changes and evaluate whether $\tau = 4$ minutes is a feasible value for determining accessibility from both geographical and network perspective. Fig. 11 describes how the region-based and population-based accessibility changed as $\tau$ increased. Accessibility for all ESS types significantly increase for small $\tau$, then reaches saturation.

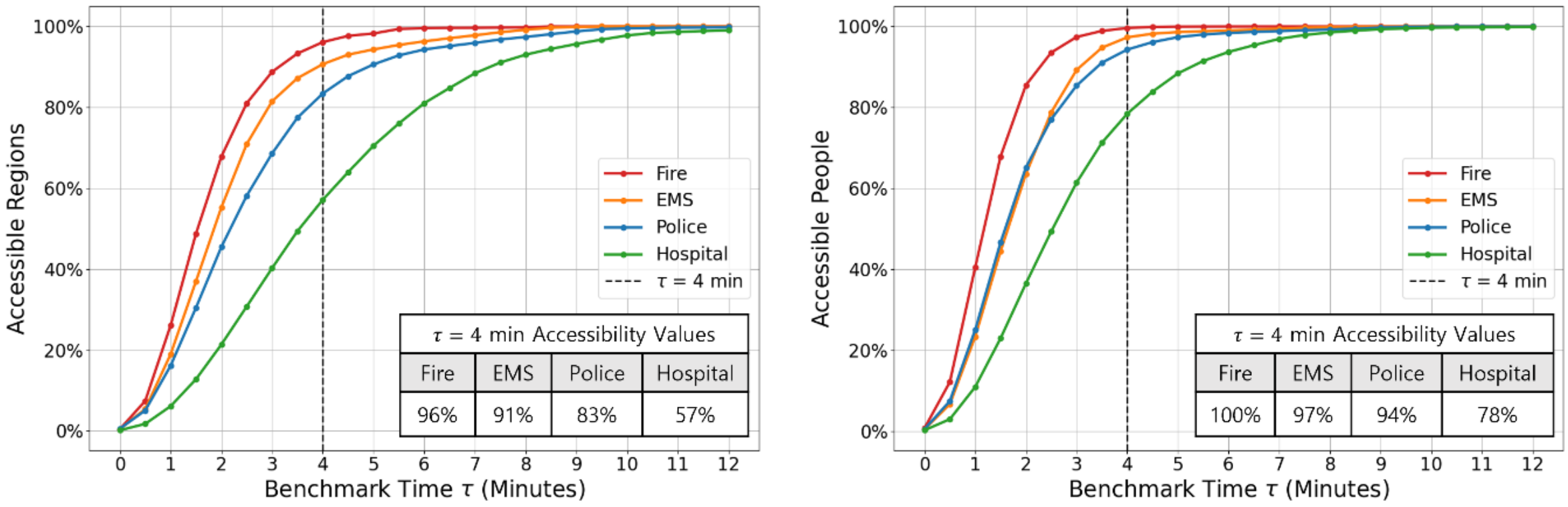

Fig. 11A: ESS accessibility based on areas covered. Fig. 11B: ESS accessibility based on people covered.

Fig. 11: Description of how accessibility of NYC changes as a function of the benchmark time $\tau$. The benchmark used to determine accessibility through the analysis ($\tau = 4$ minutes) is marked by the vertical black dashed line. Comparing the accessibility based on regions (Fig. 11A) and people (Fig. 11B), weighting the regions by population yielded a better accessibility with the same $\tau$.

For both region-based and population-based accessibilities, all services, except for Hospital, start to level off around $\tau = 4$ minutes. Also, more than 83% of the regions and 94% of populations were accessible to Fire, EMS, and Police. Even though Hospital covers 57% of the regions, it covers 78% of the population at $\tau = 4$ minutes. It is difficult to determine whether a specific percentage of coverage is ‘good enough’ for NYC, but if the goal is to provide access to emergency services for 80% of the population, then $\tau = 4$ minutes appears to be a reasonable value. At $\tau = 4$ minutes, 94% of NYC residents can access Fire, EMS, and Police, while 78% can access Hospitals. Although having a shorter travel time is better, if $\tau$ is set to less than 4 minutes then Hospital access would be too low, and it would be an infeasible goal to achieve with the current infrastructure. This leads to the conclusion that $\tau = 4$ minutes provide suitable emergency services to the majority of NYC residents. Future research that extends this methodology to other regions may find it useful to test the reasonability of a given travel time benchmark to determine the best travel time for their purposes. Policy makers may also be interested in this quantification of accessibility to better assess travel time goals for a given city.

## 4.2 Characteristics of Vulnerable Regions

In Subsection 3.1, the geographical distribution of travel times (Fig. 5) indicated that the highly accessible (dark blue) regions coincide with the vicinity of ESS sites shown in Fig. 2. This is not a surprising result because the physically closer a region is to an ESS site, the faster first responders can respond to. However, this vulnerability check was sector-agnostic. Therefore, being close to one type of ESS site does not guarantee better access to other sites. Since vulnerability depends on access to multiple types of ESS, the investigation focused on the features of regions that lack accessibility to multiple services. This investigation focused on the 5 well-defined vulnerable regions identified in Fig. 7. Observations reveal that many of these vulnerable regions have bottleneck geography and low population density.

- **Geography:** By observing the geographical locations of the five vulnerable regions in Fig. 7A-E, one can explain many of the vulnerabilities by their locations. Roosevelt Island (Fig. 7A) and City Island (Fig. 7C) are geographically vulnerable since there is only one road leading to those regions: Roosevelt Island Bridge and The City Island Bridge, respectively. Regions in Jamaica Bay (Fig. 7E) such as Rockaway Peninsula in the south and Broad Channel in the middle are also only accessible by one or two roads such as Rockaway Point Blvd, Marine Parkway Bridge, and Cross Bay Blvd. Also, these regions are too small to have all the types of ESS sites on their own. This forces some first responders inevitably go around these regions to use the only entry and exit point, causing the travel times to increase.

- **Population Density:** With the exception of Roosevelt Island, which has a high population density of 51,000 people per square mile referring to Table 2, many of these regions have low population density. Northern Queens (Fig. 7B) and City Island (Fig. 7C) have roughly 10,000 to 15,000 residents per square mile, which is also low compared to their borough's population density. Jamaica Bay area (Fig. 7E) has very low population density because it includes many federal and state lands such as JFK Airport, Gateway National Recreation Area, and Jamaica Bay Wildlife Refuge, which are not inhabitable. Inhabitable regions like Breezy Point and Broad Channel have very low population density of less than 3,000 residents per square mile. Finally, Staten Island (Fig. 7E) is known for its large number of park areas and its low population density (Table 2). Despite its size, the entire island is sparsely populated compared to other boroughs of NYC.

These findings suggest that vulnerable regions are prone to occur in regions that are remote and have low population density. Given that the impact of geography is more intuitive to understand, our focus is on the relationship between population density and accessibility to ESS.

### 4.3 Improving Overall Accessibility

This case study has shown that the majority of NYC's population is well-served by the current emergency system, but vulnerable regions still exist, indicating room for improvement in response times. While various strategies can be employed to reduce travel times during emergencies, such as enhancing staff preparedness or implementing emergency vehicle-only travel lanes, the value of this study is most evident in guiding decision-making for new ESS site construction.

The relationships shown in Subsection 3.3 can serve as quantifiable metrics for policymakers when determining optimal locations for new ESS sites. Despite the potential for an additional ESS site in a densely populated area to impact more people, the actual reduction in travel time may not cost-effectively reduce the vulnerability of the entire system. On the other hand, establishing a new ESS site in a sparsely populated area can significantly reduce the number of vulnerable regions due to the power-law relationship.

Moreover, reducing travel time from 3 minutes to 2.9 minutes may not represent a significant improvement, achieving a reduction from 5 minutes to 4 minutes can potentially save lives [Blackwell 2002, Challand 2010, Pons 2005]. As a result, even though investing in sparsely populated areas may seem inefficient, it can yield substantial benefits for individuals in vulnerable regions.

## 5 Conclusions

This study introduces a network science-based method for modeling and analyzing the accessibility of existing emergency service infrastructures. Our approach aids in understanding regional variations in access to Fire, EMS, Police, and Hospital services, thereby identifying vulnerable regions or 'emergency service deserts' in NYC.

Notably, our method considers the interconnected components of ESS, whereas previous works often treated ESS as a singular system, disregarding its multifaceted nature. Furthermore, we sought to validate our model with real-world metrics.

Overall, road network topology of NYC and the distribution of ESS site locations allow access to the majority of NYC residents (95%), indicating that the current emergency service system effectively serves the city's population. However, among the identified vulnerable regions, many share similar geographic and population characteristics. These regions are often geographically vulnerable due to limited entry and exit points, and their low population densities prompted an investigation into the relationship between inaccessibility and population density.

While previous works have primarily focused on identifying vulnerable regions, our work goes beyond by not only identifying these regions but also quantitatively investigating the characteristics and suggesting effective solutions to reduce vulnerabilities. By examining the relationship between travel times and ESS site density, we discovered a power-law relationship. This finding led us to identify Staten Island, the most sparsely populated and vulnerable borough, as having the potential to reduce vulnerability significantly with the addition of a few new ESS sites. In contrast, other boroughs are already saturated with ESS sites, and adding more has minimal and cost-inefficient impacts on reducing vulnerability. Therefore, we recommend allocating investments in less densely populated and vulnerable areas on the 'forgotten' side of NYC, such as Staten Island.

One benefit of this methodology is its simplicity, leveraging open-source data for network construction and analyzing key relationships between variables. Our methodology's flexibility and simplicity make it applicable to any region of interest. However, it is important to acknowledge some limitations, such as assuming no capacity constraints on ESS sites and not distinguishing between different services within the same sector. While it is outside the scope of this study, future endeavors should critically examine resource allocation in emergency responses. While our study presents results under normal operating conditions, our methodology can be adapted for dynamic network conditions when roads and ESS sites degrade due to stressors and disruptions. We envision adapting our methodology for resilience quantification during disruption scenarios, utilizing dynamic network models to understand how accessibility changes under different scenarios, such as Hurricane Sandy in 2012 [Aerts 2013].

# 6 Acknowledgements

We thank Stephanie Galaitsi for helpful feedback and reviewing the document.

**Funding:** This work was supported by the US Army Corps of Engineers, Engineer Research and Development Center FLEX program on compounding threats.

The views and opinions expressed in this article are those of the individual authors and not those of the US Army or other sponsor organizations.

# Supplementary Information

## 1. Definitions of Emergency Services Sector (ESS) and Locations

According to the Cybersecurity & Infrastructure Security Agency, the Emergency Services Sector is a community of trained personnel and resources that provide prevention, preparedness, response, and recovery services during both day-to-day operations and incident response [CISA 2023]. Among the five distinct disciplines composing the ESS (Fire and Rescue Services, Emergency Medical Services (EMS), Law Enforcement, Emergency Management, and Public Works), our study focused on three of them: Fire and Rescue Services, Emergency Medical Services, Law Enforcement, along with an additional service, the Emergency Departments.

The role of each service and their respective sites are described below.

- **Fire and Rescue Services**: They are referred to as Fire in the paper. They include both structural and wildland firefighting technical rescue services [CISA 2022]. Although these services often provide EMS, our study only considers fire- and rescue-related services under this sector. The representation of this sector includes fire stations and fire houses equipped with firefighting personnel and equipment for firefighting and rescue services [USGS 2023]. The data [USGS 2020] includes manned fire stations and buildings from which a fire response occurs. Fire training facilities or firefighting equipment storages without an active fire station were excluded. The data includes both full-time and volunteer fire stations, though some volunteer fire stations are missing. There are 481 Fire stations within the buffer boundary.

- **Emergency Medical Services**: They are referred to as EMS in the paper. They provide prehospital emergency medical care, with medically trained staff dispatched to administer emergency medical treatment at the scene and transport emergency patients [CISA 2022]. The level of medical service provided, whether basic or advanced, was not considered for this study. The representation of this sector involves EMS stations, which may or may not house an ambulance [USGS 2023]. In the data [HIFLD 2023a], EMS stations often overlap with fire stations since medically trained staff can respond using a non-ambulance vehicle, such as a fire engine. In this case, the site is considered a multi-role facility, and the same geographical location was used in both Fire and EMS analyses. There are 412 EMS stations within the buffer boundary.

- **Law Enforcement**: They are referred to as Police in the paper. Law Enforcement agencies are responsible for enforcing laws, maintaining public order, and managing public safety [CISA 2022]. They are publicly funded and employ at least one full-time or part-time sworn officer with general arrest powers [BJS 2016]. The representation of this sector includes police stations or sheriff's offices [USGS 2023]. The data [HIFLD 2021] includes all the local, state, and federal levels of law enforcement agencies, possibly with multiple divisions and offices. For this study, locations that are not actively dispatching units to the community were excluded, as they do not directly provide response and recovery service to the public. Excluded locations were correctional facilities, juvenile centers, and numerous specific divisions such as intelligence, internal affairs, financial, auto impounds, and many more. High school police were excluded, but university campus police were included in the analysis. Out of 331 law enforcement locations in the buffer boundary, 74 of them were excluded.

- **Emergency Departments**: They are referred to as Hospital to avoid confusion with EMS in the paper. Emergency departments are hospital facilities staffed 24 hours a day, 7 days a week, and provide unscheduled outpatient services to patients whose condition requires immediate care [CDC 2022]. Often, this service is interconnected with EMS when paramedics need to transport patients requiring more advanced medical care. For the purpose of emergency service analysis, our focus was on hospitals with emergency centers and trauma centers. We assumed every hospital listed in the data [HIFLD 2023b] had an emergency center, but psychiatric hospitals, children's hospitals, and rehabilitation centers were excluded since they focus on specific groups of people. Out of 114 hospital locations in the buffer boundary, 23 of them were excluded.

## 2. Relevant Literature

This section summarizes relevant literature by comparing and contrasting methods and results in SI Table 1.

1. The "Literature" column indicates the published findings from journals listed in the References.
2. The "Analyzed Region" column indicates the geographical scope of the studies, spanning from individual cities to entire countries. It also indicates whether a buffer zone was considered around the study area to remove boundary effects. For studies conducted on entire countries or islands, the buffer zone's impact is irrelevant and thus marked as "N/A".
3. The "Analysis Unit" column indicates the smallest geographical unit considered while calculating inaccessible regions and affected populations. It also indicates the total number of these units analyzed in the study. In cases where the paper does not explicitly state the total number of units examined, it is indicated as "Unknown".
4. The "Analyzed Population" column indicates the total number of individuals included in the analysis. If the study solely examines inaccessible regions without considering population, it is indicated as "N/A". In cases where the study analyzes affected residents but does not explicitly state the total population, it is indicated as "Unknown".
5. The "Emergency Service Type" column indicates the various types of emergency services examined in the study regarding inaccessibility. It also indicates the number of sites considered for each service type. If the paper does not explicitly state the total number of sites, it is indicated as "Unknown".
6. The "Accessibility Based On" column indicates how the methodology employed by the study determines accessibility. While many studies used travel time or distance to the nearest site, some studies used additional metrics such as bed capacity.
7. The "Disruption" column indicates whether the study considered the impact of disruptions on accessibility. If the study only examined accessibility during normal times, it is indicated as "N/A".

The geographical scope of the reviewed studies ranged from a single city level [Green 2017, Silalahi 2020, Rohr 2020, Zimmerman 2023] to a regional level [Sohn 2006, Delamater 2012, Orowhigo 2021, Khazi-Syed 2023], and to a national level [Tansley 2015, Jeon 2018, Kim 2021, Schuster 2024], with the smallest unit of analysis ranging from road intersections and census areas to municipalities and cities. Many of the studies focused on access to a single emergency system, often the healthcare system (such as emergency rooms, hospitals, clinics, or other health institutions) [Schuster 2024, Khazi-Syed 2023, Orowhigo 2021, Klipper 2021, Silalahi 2020, Tansley 2015, Delamater 2012]. However, access to Fire & Rescue, and Emergency Medical Service [Kim 2021, Rohr 2020, Jeon 2018] or Police [Eisheikh 2022, Stassen 2020] were also discussed.

Among the 14 reviewed literatures, we were particularly interested in three of them that studied access to more than one emergency services. Specifically, [Green 2017] analyzed access to Fire and EMS in Leicester, UK, based on the travel times of first responders. They identified regions vulnerable to floods based on historic flood data. Limitations include not considering affected population and a small number of fire stations (6 stations) and EMS stations (5 stations) in the region. On the other hand, [Jeon 2018] studied access to Fire and Ambulance services (fire, rescue, and emergency medical services are provided by the same organization in Korea) and Emergency Rooms for all municipalities of South Korea. It was an extensive study of emergency service accessibility for 37,683 municipalities and 51.5 million people to 509 Emergency Rooms and 1,820 Fire and Ambulance stations across South Korea. Similarly, [Kim 2021] studied changes in accessibility of municipalities across South Korea to Emergency Rooms, Fire and Ambulance services, and Public Clinics under different traffic conditions. They found that physically long distance to emergency services increased vulnerabilities in rural areas, while traffic during rush hours increased vulnerabilities in urban areas.

SI Table 1. Summary of Emergency Service Accessibility Literature

| | Literature | Analyzed Region (buffer considered?) | Analysis Unit (amount) | Analyzed Population | Emergency Service Type (Number of Sites) | Accessibility Based On | Disruption |
|---|---|---|---|---|---|---|---|
| 1 | This Study | New York City, NY, USA (Yes) | Road Intersections (55,329) | 8.4 million | Fire (481), EMS (421), Police (257), Emergency Room (91) | Travel Time | (Future Works) |
| 2 | Schuster 2024 | Austria (N/A) | Municipalities (Unknown) | Unknown | Hospital (Unknown) | Travel Time | Random Failures |
| 3 | Zimmerman 2023 | New York City, NY, USA (No) | Road Intersections (61,558) | N/A | Emergency Services (279) | Travel Time | Flood, Bridge Failure |
| 4 | Khazi-Syed 2023 | East Java & Papua, Indonesia (N/A), NCR & BARMM, Philippines (No) | Square Cells 2.3 km x 2.3 km (Unknown) | Unknown | Hospital + Clinic (Unknown) | Travel Time | N/A |
| 5 | Eisheikh 2022 | Al-Aziziyah, Jeddah, Saudi Arabia | Road Intersections (Unknown) | N/A | Police (2) | Travel Time | N/A |
| 6 | Orowhigo 2021 | Delta State, Nigeria (No) | Major Cities (50) | N/A | Health Institutions (1710) | Distance | N/A |
| 7 | Klipper 2021 | Jakarta, Indonesia (No) | Road Intersections (145,406) | 10.7 million | Emergency Room (221), Clinic (1516) | Travel Time, Bed Capacity | Flood |
| 8 | Kim 2021 | South Korea (N/A) | Municipalities (Unknown) | N/A | Emergency Room (509), Fire + EMS (1702), Clinic (3156) | Travel Time | Rush Hour |
| 9 | Rohr 2020 | Cologne, Germany (No) | Hexagonal Grid (Unknown) | N/A | Fire (Unknown) | Travel Time | N/A |
| 10 | Silalahi 2020 | Jakarta, Indonesia (No) | Villages (261) | 2637 (COVID-19 Infected) | COVID-19 Hospital (8) | Travel Time, Distance | N/A |
| 11 | Stassen 2019 | Sweden (N/A) | Square Cells 1 km x 1 km (Unknown) | 10 million | Police (95) | Travel Time | N/A |
| 12 | Jeon 2018 | South Korea (N/A) | Municipalities (37,683) | 51.5 million | Emergency Room (509), Fire + EMS (1820) | Travel Time | N/A |
| 13 | Green 2017 | Leicester, UK (No) | Road Intersections (Unknown) | N/A | Fire (6), EMS (5) | Travel Time | Flood |
| 14 | Tansley 2015 | Namibia (N/A), Haiti (N/A) | Census Area (Unknown) | 2.3 million<br>10.2 million | Hospital (410)<br>Hospital (907) | Distance | N/A |
| 15 | Delamater 2012 | Michigan, USA (No) | Square Cells 1 km x 1 km (Unknown) | 9.9 million | Emergency Room (137) | Travel Time | N/A |